\documentclass[lettersize,journal]{IEEEtran}
\usepackage{amsmath,amsfonts}
\usepackage[noend]{algpseudocode}
\usepackage{array}
\usepackage{textcomp}
\usepackage{stfloats}
\usepackage{url}
\usepackage[hidelinks,colorlinks=true,linkcolor=blue,citecolor=blue,urlcolor=black]{hyperref}
\usepackage{acronym}
\usepackage{url}
\usepackage{array}
\usepackage{tabularray}
\usepackage{verbatim}
\usepackage{subfigure}
\usepackage{amssymb}
\usepackage{graphicx}
\usepackage[dvipsnames]{xcolor}
\usepackage{cite}
\usepackage{xfrac}
\usepackage{multirow}
\usepackage{multicol}
\usepackage{matlab-prettifier}
\usepackage{mathtools}

\input{AcronymList}
\usepackage[linesnumbered,ruled,vlined]{algorithm2e}

\SetCommentSty{mycommfont}

\begin{document}
\title{Adaptive Phase-Shifted Pilot Design for Uplink Multiple Access in ISAC Systems}

\author{Ahmet Sacid S\"{u}mer, Ebubekir Memi\c{s}o\u{g}lu and H\"{u}seyin Arslan,~\IEEEmembership{Fellow,~IEEE,}
\thanks{The authors are with the Department of Electrical and Electronics Engineering, Istanbul Medipol University, Istanbul, 34810, Turkey (e-mail: ahmet.sumer@std.medipol.edu.tr; ememisoglu@medipol.edu.tr; huseyinarslan@medipol.edu.tr). \\
}
}


\maketitle
\begin{abstract}
In uplink integrated sensing and communication (ISAC) systems, pilot signal design is crucial for enabling accurate channel estimation and reliable radar sensing. In orthogonal frequency-division multiple access (OFDMA)-based frameworks, conventional pilot allocation schemes face a trade-off between spectral efficiency (SE) and sensing performance. Interleaved pilots improve user equipment (UE) multiplexing through sparse allocation but reduce the maximum unambiguous range. Conversely, orthogonal block-based pilots reduce range ambiguity but degrade sensing resolution due to limited delay granularity.
To address this trade-off, the phase-shifted ISAC (PS-ISAC) scheme was recently proposed for uplink multiple access in ISAC systems. However, PS-ISAC suffers from spectral inefficiency due to the fixed cyclic prefix (CP) constraints. To overcome these limitations, we propose adaptive phase-shifted-ISAC (APS-ISAC), an enhanced pilot scheme that employs an overlapped block-pilot structure with UE-specific phase shifts determined by each UE’s maximum excess delay. This design enables UEs to share the same time-frequency resources while preserving separable and contiguous channel impulse responses (CIRs) at the base station (BS).
Simulation results show that APS-ISAC significantly outperforms conventional pilot allocation methods in terms of SE, approximately doubling the number of multiplexed UEs. It also achieves lower mean square error (MSE) under power constraints with reduced complexity. Furthermore, it yields maximum range resolution and unambiguous sensing performance.
These results establish APS-ISAC as a scalable, spectrally efficient, ambiguity-resilient, and low-complexity pilot design paradigm for future uplink ISAC systems.
\end{abstract}


\section{Introduction}
\IEEEPARstart{I}{ntegrated} sensing and communication (ISAC) is emerging as a key enabling technology for next-generation wireless networks, especially in \ac{6G} systems, where integration of high-quality connectivity and robust situational awareness is essential~\cite{liu2022integrated, yazar20206g, 11006508}. 
By jointly leveraging spectral, hardware, and signal resources, ISAC promises gains in \ac{SE}, energy efficiency, and hardware reuse. 
Such capabilities are crucial for emerging applications, including \ac{V2X} communications and industrial \ac{IoT}~\cite{liu2024next,9585321,11060909}. 
In light of spectrum scarcity and the convergence in waveform design, radar frequency bands have emerged as promising candidates for communication reuse, motivating unified waveform design \cite{10458884}.

\Ac{OFDM} is considered a promising waveform for ISAC, given its maturity in existing wireless standards (e.g., \ac{5G}, WiFi), high \ac{SE}, robustness against multipath fading, and signal design flexibility~\cite{keskin2025fundamental, nataraja2024integrated}. Among various ISAC architectures, the monostatic approach has gained popularity due to its capability to leverage transmitted signals fully for radar sensing and mitigate synchronization complexities associated with bistatic or multistatic setups~\cite{pucci2022system, keskin2023monostatic}. Particularly, \ac{OFDM}-based monostatic sensing exploits pilots to estimate range and Doppler profiles, benefiting from their strong autocorrelation and known structure~\cite{kumari2017ieee, Wifi_sensing, 5G_PRS}.

Despite these advantages, designing efficient multiple access schemes for ISAC systems remains challenging, requiring a careful balance between sensing accuracy, communication throughput, and interference mitigation~\cite{liu2022evolution}. Conventional multiple access methods such as time-, frequency-, or code-division multiplexing each introduce trade-offs: Frequency division maintains unambiguous velocity estimation but severely constrains range; time division ensures range resolution but sacrifices Doppler estimation and requires stringent synchronization; code division schemes enable simultaneous access but introduce cross-interference, impairing sensing accuracy, especially for weak targets \cite{nuss2018limitations}. Moreover, maintaining low side-lobes and ambiguity in sparse pilot configurations is crucial yet challenging in modern \ac{OFDM}-based sensing systems~\cite{mura2024optimized}.

Pilot design plays a critical role in enabling multiple access in ISAC systems, where uplink pilot reuse has been explored through overlapped interleaved structures and pilot preambles to support multi-\ac{UE} access~\cite{Ribeiro2008UplinkCE,kumari2017ieee}.
However, these methods often incur high overhead or suffer from degraded estimation under frequency-selective fading~\cite{overlapping_pilots_2005}, while largely overlooking ISAC-specific metrics such as unambiguous range, delay resolution, and efficient spectrum utilization. In \ac{OFDMA}-based frameworks such as \ac{LTE} and \ac{5G} \ac{NR}, orthogonal interleaved and block pilot patterns designed primarily for communication have been repurposed for sensing~\cite{OFDMA_survey}, but impose trade-offs: higher pilot densities improve sensing accuracy at the cost of multiplexing, while sparse allocation enhances scalability but degrades sensing performance~\cite{3gpp2025_38211,liu2024next}.

\subsection{Review of Related Studies on ISAC Pilot Design}
\label{subsec:review}
In the downlink, OFDM radar techniques utilizing dynamic subcarrier allocation~\cite{Ozkaptan_1,Ozkaptan_2} and bi-static sensing-aware \ac{RS} pattern designs~\cite{zhang2024ofdm} have been proposed. Energy-efficient joint \ac{RS}-power optimization for \ac{V2X} ISAC~\cite{zhao2023reference} and the use of \ac{5G} \ac{PRS} for radar~\cite{5G_PRS} further demonstrate the potential of structured pilot reuse. Hybrid \ac{PRS}-\ac{DMRS} schemes, such as interleaved pilots~\cite{ni2024frequency}, offer moderate gains in unambiguous range but are constrained by the complexity of compressed sensing~\cite{10561589}. Recent works also explore overlapping pilot-sensing sequences~\cite{memisoglu2023csi,demir2023csi} and mutual information–driven joint designs~\cite{bazzi2025mutual}, though most rely on fixed interleaved patterns, iterative estimation, or quantization-sensitive recovery.

To improve unambiguous range, coprime pilot design~\cite{Coprime_ISAC} and non-equidistant interleaving~\cite{nuss2020frequency,hakobyan2019ofdm} have been studied, but these approaches often increase implementation complexity. Sequential and hybrid schemes that combine orthogonal block and interleaved pilots~\cite{xiao2024achieving} improve sensing resolution but introduce latency and increased energy overhead. Other strategies such as staggered subcarrier offsets~\cite{zhang2024staggered}, \ac{CRB}-optimized time-frequency allocation~\cite{mura2024waveform}, and delay-Doppler waveform superposition~\cite{tagliaferri2023integrated} involve trade-offs among complexity, sidelobe suppression, and standard compatibility. However, none of these approaches addresses scalable uplink pilot reuse in multi-\ac{UE} scenarios.

Despite substantial research, scalable, and efficient pilot allocation strategies for uplink multi-UE ISAC remain underdeveloped. Existing methods often fail to fully exploit bandwidth, ensure delay-domain separability, or maintain low implementation complexity. In our previous work~\cite{PS-ISAC}, we introduced a \ac{CP}-based phase-shifted pilot scheme that laid the groundwork for full pilot reuse among multiple \acp{UE}. However, several critical limitations remained unresolved.
First, the \ac{SE} of \ac{PS-ISAC} matches that of \ac{CI-ISAC}, since \ac{PS-ISAC} is constrained by the \ac{CP} ratio, whereas \ac{CI-ISAC} is limited by the pilot ratio, resulting in no improvement in \ac{UE} multiplexing. Second, the sensing model was restricted to \ac{CSI}-based approaches and did not account for radar sensing with multiple targets. Third, the trade-off between range ambiguity and resolution remained theoretically uncharacterized.

\subsection{Motivation and Key Contributions}
The aforementioned challenges in Subsection~\ref{subsec:review} motivate an ISAC-specific multiple access solution that (i) supports simultaneous access by multiple \acp{UE} over shared spectral resources, (ii) maximizes pilot reuse while maintaining estimation accuracy and low-complexity, and (iii) enhances radar resolution and extends the maximum unambiguous range under practical spectral and power constraints.

To this end, we introduce an adaptive phase-shift pilot allocation strategy termed \ac{APS-ISAC} designed for uplink multi-UE ISAC systems. Building upon the foundation laid in~\cite{PS-ISAC}, APS-ISAC employs an overlapped block-pilot structure with UE-specific phase shifts, dynamically determined by each UE’s maximum excess delay. This design enables scalable pilot reuse, ensures reliable \ac{CIR} separability, and enhances radar sensing accuracy under both \ac{CSI}-based and radar-based sensing models. The key contributions of this work are summarized as follows:

\begin{itemize}

\item \textbf{Adaptive Phase-Shifted Pilot Design for Multi-UE ISAC Uplink:} We propose \ac{APS-ISAC}, a novel adaptive phase-shifted pilot allocation scheme for uplink multiple access in ISAC systems. By combining full-band pilot transmission with \ac{UE}-specific phase shifts determined by the maximum excess delay, \ac{APS-ISAC} enables precise and scalable separation of contiguous \acp{CIR} at the \ac{BS}, thereby fully utilizing time-domain resources for CIR separation. The design supports both CSI-based sensing across multiple \acp{UE} and monostatic radar-based sensing for multiple targets within a unified uplink framework.

\item \textbf{Scalable Pilot Reuse and Enhanced UE Multiplexing with Low-Complexity:} \ac{APS-ISAC} supports full pilot reuse and significantly improves \ac{UE} multiplexing. For a given channel model, it approximately doubles the number of simultaneously supported \acp{UE} compared to \ac{PS-ISAC} and \ac{CI-ISAC}, owing to its ability to efficiently separate contiguous \acp{CIR}. Furthermore, it achieves lower per-\ac{UE} computational complexity at the \ac{BS} with negligible control signaling overhead. We also analyze how the multiplexing gain scales with delay spread statistics and derive an upper bound on the number of multiplexed \acp{UE} under frequency-flat channels.

\item \textbf{Joint Range Resolution and Ambiguity Optimization:} We formally analyze the trade-off between range resolution and unambiguous range for both conventional and proposed pilot designs, and demonstrate that \ac{APS-ISAC} effectively resolves this trade-off via  full-band pilot allocation. This is further validated via radar-based range–velocity simulations.

\end{itemize}

\subsection{Outline of This Paper}
The remainder of this paper is organized as follows. Section~\ref{sec:system_model} describes the uplink OFDM-based multiple access framework for ISAC \acp{UE} performing monostatic sensing, covering both the communication and radar channel models, the BS-side operation for CSI-based sensing, and the UE-side operation for radar-based monostatic sensing. Section~\ref{sec:prop_method} introduces the proposed adaptive phase-shift pilot design, emphasizing its sensing capabilities, control signaling overhead, computational complexity, and the inherent trade-off between maximum unambiguous range and range resolution in pilot design for ISAC systems. Section~\ref{sec:sim_results} presents detailed simulation results that validate the performance advantages of the proposed scheme. Finally, Section~\ref{sec:conclusion} concludes the paper and discusses potential directions for future research.
\subsection{Notations}
Bold uppercase and lowercase letters denote matrices and vectors, respectively (e.g., $\mathbf{X}$, $\mathbf{x}$), while non-bold letters (e.g., $X$, $x$) represent scalars. Calligraphic symbols such as $\mathcal{X}$ denote sets, and $\mathbb{N}$ represents the set of natural numbers. The Dirac delta function and the expectation operator are denoted by $\delta(\cdot)$ and $\mathbb{E}[\cdot]$, respectively. $(\cdot)^{i}$ indicates the scheme index, where \( i \in \{\text{aps}, \text{ps}, \text{ci}, \text{cb}\} \) corresponds to \ac{APS-ISAC}, \ac{PS-ISAC}, \ac{CI-ISAC}, and \ac{CB-ISAC}, respectively. As a general convention, the notations \( \bar{(\cdot)} \), \( \hat{(\cdot)} \), and \( \tilde{(\cdot)} \) represent variables related to the transmit side, the receive side of the ISAC \ac{UE}, and the ISAC BS, respectively. The ceiling and absolute value operations are represented by $\lceil \cdot \rceil$ and $|\cdot|$. We use $\mathcal{CN}(\mu, \sigma^2)$ and $\mathcal{N}(\mu, \sigma^2)$ to denote circularly symmetric complex and real Gaussian random variables with mean $\mu$ and variance $\sigma^2$, respectively, and $\Gamma(\alpha, \theta)$ to represent the Gamma distribution with shape parameter $\alpha$ and scale parameter $\theta$.

\begin{figure*}[t]
    \centering
    \hspace*{-0.06\textwidth}
    \includegraphics[width=1.14\textwidth]{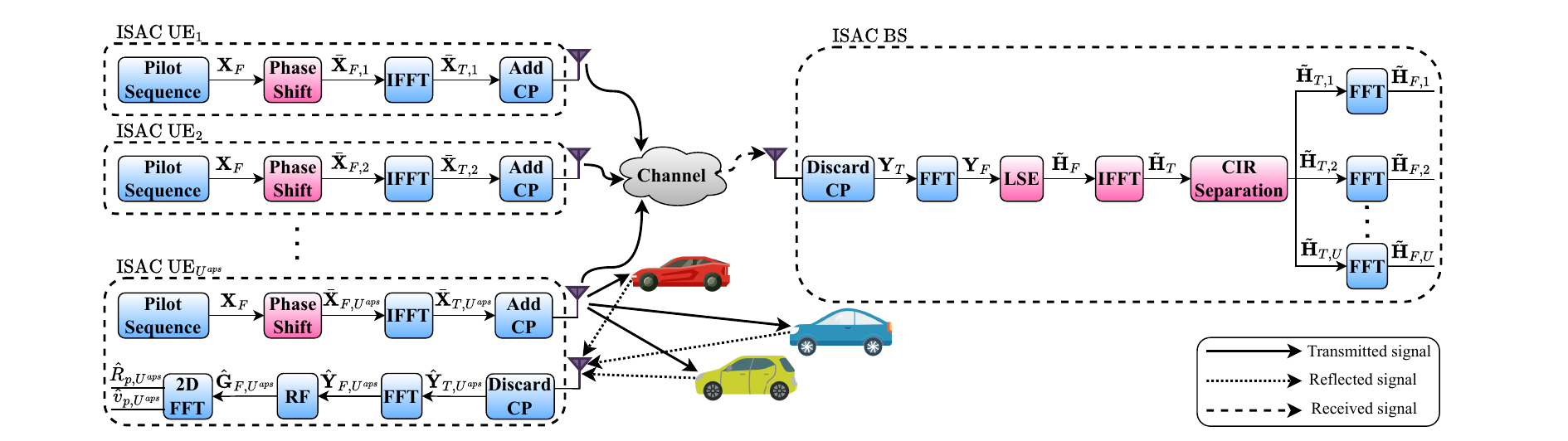}
    \caption{Block diagram of the proposed uplink ISAC multiple-access scheme employing \ac{UE}-specific phase shifts.}
    \label{fig:system_model}
\end{figure*}

\section{System Model}
\label{sec:system_model}

As illustrated in Fig.~\ref{fig:system_model},\footnote{Blocks that differ from those in \ac{PS-ISAC} and \ac{CI-ISAC} are highlighted in pink.}  we consider an uplink multi-\ac{UE} ISAC scenario where \( U^{i} \) single-antenna ISAC \acp{UE} simultaneously transmit pilot-bearing uplink \ac{OFDM} signals to a centralized ISAC BS in the presence of \( P \) radar targets. The system adopts a \ac{MAC} topology with a monostatic configuration, in which each ISAC \ac{UE} performs radar sensing by exploiting reflections from its own transmitted pilot signals. This setup enables simultaneous uplink channel estimation and target parameter extraction, thereby supporting vehicular and short-range applications with unified transceiver functionality.~\cite{Sturm2011,keskin_foe,10681580}.
The monostatic configuration facilitates perfect synchronization by sharing the same hardware between the transmit and receive chains for radar-based sensing.~\cite{Sturm2011}. Full-duplex operation is assumed for monostatic sensing, supported by ideal transmit-receive isolation achieved through advanced analog and digital self-interference cancellation techniques~\cite{barneto2019full,biedka2019full}.

\subsection{Transmit Signal Model}
\label{subsec:transmit_signal_model}

Each \ac{UE} embeds a known pilot signal in a block of \( M \) \ac{OFDM} symbols, each consisting of \( N \) subcarriers. Let \( \mathcal{M} = \{0, 1, \dots, M{-}1\} \) and \( \mathcal{K} = \{0, 1, \dots, N{-}1\} \) denote the index sets of OFDM symbols and subcarriers, respectively. The OFDM symbol duration is denoted by \( T \), which yields a subcarrier spacing of \( \Delta f = 1/T \). To enable joint communication channel estimation and radar target sensing, each \ac{UE} \( u \in \{1, \dots, U^{i}\} \) generates a block-pilot matrix \( \mathbf{X}_F \in \mathbb{C}^{N \times M} \), populated by a known pseudo-random pilot sequence. These pilots are modified using \ac{UE}-specific adaptive phase shifts, determined by the maximum excess delay between each ISAC \ac{UE} and the ISAC BS.

The time-domain OFDM signal at the \( u \)-th \ac{UE}, corresponding to the \( n \)-th sample of the \( m \)-th OFDM symbol, is obtained via the \ac{IFFT} as
\begin{equation}
\label{eq:time_domain_transmit_signal}
\bar{X}_{T,u}(n,m) \!=\! \frac{1}{\sqrt{N}}\! \sum_{k=0}^{N-1}  \! \bar{X}_{F,u}(k,m) \, e^{j2 \pi k  n / N}, \!\!\!\!\! \quad n \!\in\! \mathcal{K}, \, m \!\in\! \mathcal{M},
\end{equation}
where \( \bar{X}_{F,u}(k,m) \) denotes the frequency-domain pilot symbol of the \( u \)-th \ac{UE} at the \( m \)-th OFDM symbol and \( k \)-th subcarrier after adaptive phase shifting.

In ISAC OFDM waveforms, the CP mitigates \ac{ISI} and preserves subcarrier orthogonality. Additionally, the maximum detectable delay corresponding to the round-trip delay of the farthest target is constrained by the CP duration \( T_{cp} \), associated with an \( N_{cp} \)-point CP. To accommodate both communication and radar requirements, the CP length must satisfy

\begin{equation}
N_{cp} \geq \max \left\{ L_{\max}, \left\lceil \frac{2 f_s R_{\max}}{c} \right\rceil \right\},
\end{equation}
where \( f_s \) is the sampling frequency and \( c \) is the speed of light. The first term, \( L_{\max} = \max_{1 \leq u \leq U^{i}} L_u \), denotes the maximum number of channel taps among all \acp{UE}, where \( L_u \) is the number of resolvable channel taps for the \( u \)-th \ac{UE}. \( L_{\max} \) and \( L_u \) reflect the maximum excess delay spread of the channel, with larger values indicating a greater number of resolvable taps. Each UE may have a different value of \( L_u \), determined by its specific channel characteristics~\cite{zhang2011joint}. The second term accounts for the maximum round-trip delay among all targets, and it is associated with \( R_{\max} \), which denotes the maximum range from any target to all \acp{UE}, and it is given by
\begin{equation}
R_{\max} = \max_{\substack{1 \leq u \leq U^{i} \\ 1 \leq p \leq P}} R_{p,u}
\end{equation}
where \( R_{p,u} \) is the range between the \( p \)-th target and the \( u \)-th \ac{UE}. 
After CP addition, the signal is passed through a \ac{DAC} for transmission.

\subsection{Channel Models}
\label{subsec:channel_models}

In the considered uplink ISAC scenario, we model the radar sensing and communication channels separately to capture their distinct physical characteristics under practical vehicular conditions. The radar channel accounts for round-trip propagation effects from dynamic targets, while the communication channel models single-trip wireless propagation between each ISAC \ac{UE} and the ISAC BS.

\subsubsection{Radar Sensing Channel}
\label{subsec:radar_sensing_channel}

We adopt a linear time-varying radar channel model, where each ISAC \ac{UE} observes \( P \) distinct point targets (e.g., vehicles), each modeled as a single-point scatterer with a constant \ac{RCS} over a short \ac{CPI}~\cite{kumari2017ieee}. The baseband-equivalent impulse response at the \( u \)-th \ac{UE} is given by~\cite{luong2021radio}
\begin{equation}
G_{T,u}(t, \tau) = \sum_{p=1}^{P} \alpha_{p,u} \, \delta(t - \tau_{p,u}) \, e^{j 2\pi \nu_{p,u} t},
\label{eq:radar_channel_time}
\end{equation}
where \( \alpha_{p,u} \sim \mathcal{CN}(0,1) \) denotes the complex reflection coefficient of the \( p \)-th target observed by the \( u \)-th ISAC \ac{UE}, which is assumed to remain constant over the \ac{CPI}. The terms \( \tau_{p,u} \) and \( \nu_{p,u} \) represent the corresponding round-trip delay and Doppler shift, respectively, and are given by
\begin{equation}
\tau_{p,u} = \frac{2 R_{p,u}}{c}, \qquad \nu_{p,u} = \frac{2 v_{p,u} f_c}{c},
\end{equation}
where \( v_{p,u} \) is the relative velocity of the \( p \)-th target with respect to the \( u \)-th \ac{UE}, and \( f_c \) is the carrier frequency. Targets are assumed quasi-static over the \ac{CPI}, ensuring that \( \tau_{p,u} \) and \( \nu_{p,u} \) remain constant across \( M \) OFDM symbols.

By projecting the radar response onto the OFDM basis via a Fourier transform over delay and sampling in slow time at \( t = mT \), the frequency-domain radar channel becomes~\cite{keskin_foe}
\begin{equation}
G_{F,u}(k, m) = \sum_{p=1}^{P} \alpha_{p,u} \, e^{-j 2\pi k \Delta f \tau_{p,u}} \, e^{j 2\pi m T \nu_{p,u}}.
\label{eq:radar_channel_freq}
\end{equation}
Since \( \Delta f \gg \nu_{p,u} \) under typical vehicular conditions, Doppler-induced \ac{ICI} is negligible~\cite{keskin2021limited}, and Doppler effects manifest primarily as phase rotations across OFDM symbols. Channel stationarity is ensured by assuming \( v_{p,u} \ll c \) for all targets. The number of targets \( P \) is assumed to be known or estimated via information-theoretic criteria such as \ac{AIC}~\cite{fishler2002detection} or \ac{MDL}~\cite{wax1989detection}.

\begin{figure*}
    \centering
    \subfigure[Conventional methods: 1) CI-ISAC, 2) CB-ISAC, and 3) PS-ISAC.]{\includegraphics[width=0.49\textwidth]{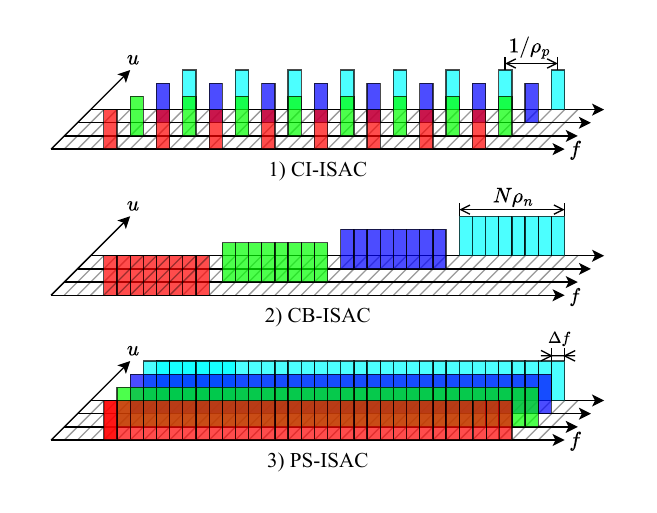}} 
    \subfigure[Proposed method: 4) APS-ISAC.]{\includegraphics[width=0.5\textwidth]{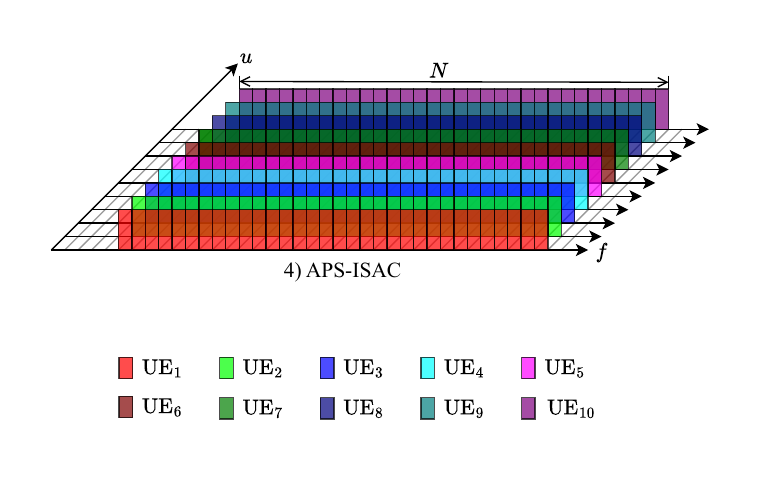}} 
    \caption{Illustration of \ac{UE}-side frequency-domain structures for both conventional and proposed pilot design methods.}
    \label{fig:tx_view}
\end{figure*}

\subsubsection{Communication Channel}
\label{subsubsec:comm_channel}

The uplink communication channel between the \( u \)-th ISAC \ac{UE} and the ISAC BS is modeled as a frequency-selective, time-invariant multipath channel with \( L_u \) taps~\cite{Sturm2011, zhao2023reference}. The baseband-equivalent time-domain impulse response is expressed as
\begin{equation}
\label{eq:comm_channel_time}
H_{T,u}(\tau) = \sum_{l=0}^{L_u-1} \beta_{u,l} \, \delta(\tau - \tau_{u,l}),
\end{equation}
where \( \beta_{u,l} \sim \mathcal{CN}(0,1) \) and \( \tau_{u,l} \) denote the complex channel gain and propagation delay of the \( l \)-th path associated with the \( u \)-th \ac{UE}, respectively. The channel is assumed to remain constant over the duration of an OFDM symbol but may vary over longer time scales. The corresponding frequency-domain channel response is given by~\cite{zhao2023reference}
\begin{equation}
\label{eq:comm_channel_freq}
H_{F,u}(k,m) = \sum_{l=0}^{L_u-1} \beta_{u,l} \, e^{-j 2\pi k \Delta f \tau_{u,l}}.
\end{equation}
The number of channel taps \( L_u \) is assumed to be known a priori, as it directly corresponds to the the maximum delay spread of the UE's communication channel, which can be practically measured~\cite{zhao2023reference, rappaport2015wideband}.

\subsection{Radar-Based Sensing}
\label{subsec:radar_BS}

Following \ac{ADC}, \ac{CP} removal, and \ac{SP} conversion, the received time-domain signal $\mathbf{\hat{Y}}_{T,u}$ at each \ac{UE} is sampled at a rate \( f_s = N \Delta f \). Applying an \( N \)-point FFT to each OFDM symbol yields the frequency-domain received signal for the \( u \)-th \ac{UE}
\begin{equation}
\begin{aligned}
&\hat{Y}_{F,u}(k, m) = G_{F,u}(k, m) \bar{X}_{F,u}(k, m) + W_{F,u}(k, m) \\ =&\! \sum_{p=1}^{P}\! \alpha_{p,u} e^{-j 2\pi k \Delta f \tau_{p,u}} \, e^{j 2\pi m T \nu_{p,u}}  \bar{X}_{F,u}(k, m)\! + \! W_{F,u}(k, m),
\end{aligned}
\label{eq:transmitter_radar_model}
\end{equation}
where \( W_{F,u}(k, m) \sim \mathcal{CN}(0, \sigma^2) \) denotes the \ac{AWGN} at the \( u \)-th \ac{UE} with variance \( \sigma^2 \).

In the proposed \ac{APS-ISAC} framework, all subcarriers are populated with known pilot symbols, enabling full-band radar response estimation. The \ac{RF} module at the \( u \)-th \ac{UE} computes the per-subcarrier response by performing element-wise division of the received symbols by the known transmitted symbols, as shown in~\cite{keskin2025fundamental}
\begin{equation}
\begin{aligned}
    & \hat{G}_{F,u}(k, m) = \frac{\hat{Y}_{F,u}(k, m)}{\bar{X}_{F,u}(k, m)} \\
    &= \sum_{p=1}^{P} \alpha_{p,u} \, e^{-j 2\pi k \Delta f \tau_{p,u}} \, e^{j 2\pi m T \nu_{p,u}} + \frac{W_{F,u}(k, m)}{\bar{X}_{F,u}(k, m)}.
\end{aligned}
\label{eq:fullband_chan_est}
\end{equation}
The estimator in~\eqref{eq:fullband_chan_est} corresponds to the \ac{ZF} solution and can be interpreted as the outcome of a least squares formulation based on~\eqref{eq:transmitter_radar_model}, as discussed in~\cite{ozdemir2007channel}.

A two-dimensional FFT (2D-FFT) applied across \( N \) subcarriers and \( M \) OFDM symbols of \( \hat{G}_{F,u}(k, m) \) yields the delay-Doppler map for the \( u \)-th \ac{UE}~\cite{Sturm2011}
\begin{equation}
\Psi_{\mathrm{DD},u}(\mathfrak{d}, \ell) = \frac{1}{NM} \left| \sum_{m=0}^{M-1} \sum_{k=0}^{N-1} \hat{G}_{F,u}(k, m) \, e^{-j \frac{2\pi k \mathfrak{d}}{N}} \, e^{j \frac{2\pi m \ell}{M}} \right|^2,
\label{eq:rd_map}
\end{equation}
where \( \mathfrak{d} \) and \( \ell \) are the delay and Doppler bin indices, respectively. Peaks in \( \Psi_{\mathrm{DD},u}(\mathfrak{d}, \ell) \) correspond to the estimated target ranges \( \hat{R}_{p,u} \) and velocities \( \hat{v}_{p,u} \).
Although \ac{CFAR}-based detectors can be applied~\cite{7870764}, this work adopts a fixed-threshold detection method~\cite{memisoglu2022orthogonal}.

\subsection{CSI-Based Sensing}
\label{subsec:csi_isac_BS}

In the proposed uplink APS-ISAC framework, the ISAC BS jointly performs channel estimation and CSI-based sensing by processing the aggregated uplink signal from all \( U^{i} \) \acp{UE}. 
After \ac{ADC} and \ac{CP} removal, the received time-domain signal 
\( \mathbf{Y}_{T} \) is transformed into the frequency domain via an 
$N$-point FFT, yielding
\begin{equation}
\begin{aligned}
& Y_F(k,m) = \sum_{u=1}^{U^{i}} H_{F,u}(k, m) \, \bar{X}_{F,u}(k, m) + W_F(k, m) \\
         &= \sum_{u=1}^{U^{i}} \sum_{l=0}^{L_u-1} \beta_{u,l} \, e^{-j 2\pi k \Delta f \tau_{u,l}} \, \bar{X}_{F,u}(k, m) + W_F(k, m),
\end{aligned}
\label{eq:rx_freq_signal}
\end{equation}
where \( W_F(k, m) \sim \mathcal{CN}(0, \sigma^2) \) denotes the \ac{AWGN} at the ISAC BS with variance $\sigma^2$. Perfect time and frequency synchronization is assumed across all \acp{UE} \cite{memisoglu2023waveform}.

\begin{figure*}[t]
    \centering
    \subfigure[APS-ISAC.]
        {\includegraphics[width=0.32\textwidth]{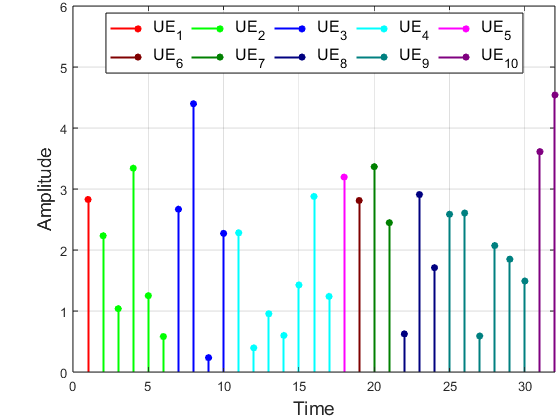}}
    \hfill
    \subfigure[PS-ISAC.]
        {\includegraphics[width=0.32\textwidth]{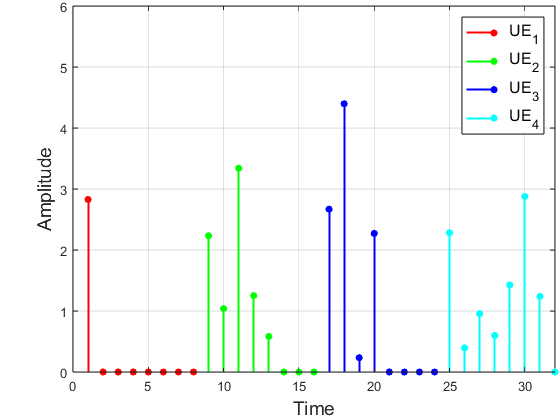}}
    \hfill
    \subfigure[CI-ISAC.]
        {\includegraphics[width=0.32\textwidth]{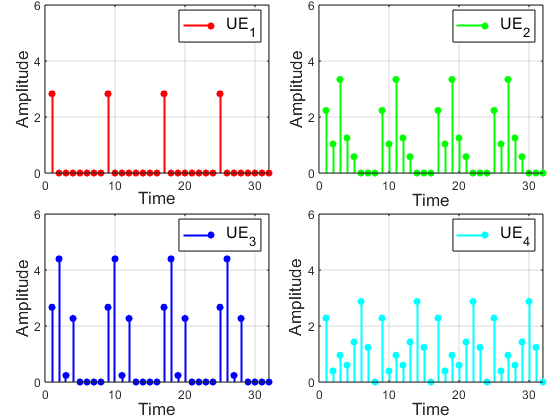}}
    \caption{Comparison of BS-side CIR structures of \acp{UE} for different pilot design methods.}
    \label{fig:CIRs_method}
\end{figure*}

\section{Proposed Adaptive Phase Shift Design}
\label{sec:prop_method}

Conventional OFDMA-based ISAC systems, such as \ac{CI-ISAC} and \ac{CB-ISAC}, rely on orthogonal interleaved or block pilot allocation strategies to maintain strict orthogonality across \acp{UE}. While this ensures reliable channel estimation, it significantly limits SE by reducing  the number of simultaneously supported \acp{UE} and degrading sensing capabilities. Specifically, \ac{CI-ISAC} suffers from a reduced maximum unambiguous range due to the widened subcarrier spacing between pilot-bearing subcarriers~\cite{hakobyan2019ofdm}, whereas \ac{CB-ISAC} experiences degradation in range resolution as a result of utilizing only a subset of the available bandwidth~\cite{OFDMA_survey}. PS-ISAC addresses these sensing limitations by allocating the full spectrum to each \ac{UE}, but still faces scalability constraints imposed by the fixed \ac{CP} ratio~\cite{PS-ISAC}. To overcome these limitations, we propose a novel pilot design framework, referred to as \ac{APS-ISAC}, in which adaptive, \ac{UE}-specific phase shifts are applied in the frequency domain to pilot symbols This enables efficient \ac{CIR} separation at the ISAC BS, thereby supporting a larger number of \acp{UE} without sacrificing spectral or sensing performance.

As illustrated in Fig.~\ref{fig:tx_view}, the pilot ratio is configured as \( \rho_p = 1/4 \), with \( N = 32 \) subcarriers and \( N_{cp} = 8 \), yielding a CP ratio of \( \rho_{cp} = 1/4 \) to ensure a fair comparison. We also define \( \rho_n = 1/4 \), representing the ratio of \( N \) to the number of subcarriers allocated to each \ac{UE} in \ac{CB-ISAC}. Fig.~\ref{fig:tx_view}(a) shows the pilot allocation in \ac{CI-ISAC}, \ac{CB-ISAC}, and PS-ISAC, while Fig.~\ref{fig:tx_view}(b) depicts the proposed APS-ISAC structure. The maximum number of \acp{UE} supported by conventional schemes is determined by the specified parameters given as
\begin{equation}
\label{eq:conv_users}
     U^{ci} = \frac{1}{\rho_p} = 4, \quad U^{cb} = \frac{1}{\rho_n} = 4, \quad U^{ps} = \frac{1}{\rho_{cp}} = 4,
\end{equation}
where \( U^{ci} \), \( U^{cb} \), and \( U^{ps} \) represent the number of \acp{UE} supported in \ac{CI-ISAC}, \ac{CB-ISAC}, and \ac{PS-ISAC}, respectively.
In contrast, APS-ISAC enables flexible and scalable \ac{UE} multiplexing by allowing simultaneous full-band pilot transmission while leveraging time-domain separation to extract contiguous CIRs. The maximum number of supported \acp{UE} is defined as
\begin{equation}
\label{eq:prop_users}
    U^{aps} = \max \left\{ U \in \mathbb{N} \,\middle|\, \sum_{u=1}^{U} L_u \leq N \right\},
\end{equation}
where \ac{UE}-specific phase shifts ensure that the resulting CIRs do not overlap at the ISAC BS. This approach enables robust channel separation and supports interference-free uplink ISAC sensing in practical Rayleigh fading environments, where the number of channel taps is assumed to follow a truncated normal distribution within the range \( [1, N_{cp} - 1] \). For instance, under such conditions, a single channel realization may support the simultaneous multiplexing of up to 10 \acp{UE} as shown in Fig.~\ref{fig:tx_view}, demonstrating the scalability of the proposed design.

\subsection{Phase Shift Mapping at ISAC \ac{UE}}
\label{subsec:phase_shift_mapping}

Prior to transmission, each \ac{UE} applies a unique cyclic phase-shift function across the subcarriers, as illustrated in Fig.~\ref{fig:system_model}. The frequency-domain pilot signal after phase-shift mapping for the $u$-th UE is given by
\begin{equation}
\bar{X}_{F,u}(k,m) = \phi_u(k,m) X_F(k,m),
\label{eq:aps_phase_applied}
\end{equation}
where \( \phi_u(k,m) \) is the \ac{UE}-specific phase-shift term defined as
\begin{equation}
\phi_u(k,m) = e^{-j 2\pi m k n_u / N}, \quad n_u = \sum_{i=0}^{u-1} L_i,
\label{eq:aps_phase_term}
\end{equation}
where \( n_u \) denoting the effective time-domain offset for the \( u \)-th \ac{UE}’s channel. This offset is computed based on the cumulative number of channel taps of preceding \acp{UE}, ensuring non-overlapping \acp{CIR} in the time domain. As a result, post-IFFT separation of \acp{CIR} at the ISAC BS is achieved without requiring orthogonal pilot allocation.

\subsection{CIR Separation at ISAC BS}
\label{subsec:cir_separation}

To estimate the composite \ac{CIR}, the ISAC BS first computes the frequency-domain channel response by performing element-wise division of the received signal by the known pilot symbols
\begin{equation}
\Tilde{H}_{F}(k,m) = \frac{Y_F(k,m)}{X_{F}(k,m)}.
\label{eq:freq_response_estimation}
\end{equation}
An \( N \)-point \ac{IFFT} is then applied along each column to obtain the time-domain representation
\begin{equation}
\Tilde{H}_{T}(n,m) = \frac{1}{\sqrt{N}} \sum_{k=0}^{N-1} \Tilde{H}_{F}(k,m) \, e^{ j 2\pi k n /N}.
\label{eq:cir_estimation}
\end{equation}
This operation yields a superimposed CIR, where each \ac{UE}'s channel occupies a distinct, non-overlapping region determined by the phase-shift design. The ISAC BS extracts the \( u \)-th \ac{UE}'s contribution from the segment \( n \in [n_u, n_u + L_u) \) as \( \mathbf{\Tilde{H}}_{T,u} \), and reconstructs its individual \ac{CFR} via FFT as
\begin{equation}
\Tilde{H}_{F,u}(k,m) = \sqrt{N} \sum_{n=n_u}^{n_u + L_u - 1} \Tilde{H}_T(n,m) \, e^{- j 2\pi n k/N}.
\label{eq:fft_extract_cfr}
\end{equation}

Fig.~\ref{fig:CIRs_method} illustrates the time-domain separation of the \acp{CIR} belonging to each \ac{UE} at the ISAC BS, comparing the proposed APS-ISAC with \ac{PS-ISAC} and \ac{CI-ISAC}. \ac{CB-ISAC} is excluded from this comparison due to its limited delay resolution; however, it is included in Subsection~\ref{subsec:tradeoff_delay_resolution}, where the trade-off between delay resolution and maximum unambiguous delay range is analyzed. 
In APS-ISAC, \ac{UE}-specific frequency-domain phase shifts are assigned based on each user’s maximum number of channel taps, ensuring that the resulting \acp{CIR} occupy contiguous, non-overlapping segments in the time domain without relying on orthogonality. This enables reliable \ac{CIR} separation and supports scalable \ac{UE} multiplexing, allowing up to 10 \acp{UE} to share the same time-frequency resources in a typical realization of randomly distributed channel tap lengths.
Although \ac{PS-ISAC} also employs full-band pilot transmission, it does not utilize adaptive phase shifts. As a result, the \acp{CIR} are confined to non-contiguous segments spread across \( N_{cp} \), leading to inefficient time-domain utilization. On the other hand, \ac{CI-ISAC} performs separation in the frequency domain using interleaved pilots, which introduces a periodic structure in the time domain. In contrast, the block-overlapped design of APS-ISAC avoids such periodicity and achieves effective time-delay domain segmentation, supporting scalable and interference-free \ac{UE} multiplexing.

\begin{table*}[t]
\centering
\caption{Computational complexity comparison of APS-ISAC, PS-ISAC, and CI-ISAC schemes at both the \ac{UE} and BS sides.} 
\label{table:complexity}
\begin{tabular}{l|ll|ll|}
\cline{2-5}
                             & \multicolumn{2}{l|}{Transmitter Side of ISAC \Ac{UE}}                                      & \multicolumn{2}{l|}{ISAC BS}                                                                             \\ \hline
\multicolumn{1}{|l|}{\!\!\!Method\!\!\!} & \multicolumn{1}{l|}{Addition}                             & Multiplication                     & \multicolumn{1}{l|}{Addition}                                    & Multiplication                              \\ \hline
\multicolumn{1}{|l|}{APS}    & \multicolumn{1}{l|}{$\!\!U^{aps}(3N\log_{2}N \!- \!3N+4)\!\!$} & $\!\!U^{aps}(N\log_{2}N \!-\!N+4)\!\!$ & \multicolumn{1}{l|}{$\!\!(U^{aps}+2)(N\log_{2}N \!-\!3N+4)+ 2N\!\!$} & $\!\!(U^{aps}+2)(N\log_{2}N \!-\!3N+4)+ 2N\!\!$ \\ \hline
\multicolumn{1}{|l|}{PS}     & \multicolumn{1}{l|}{$\!\!U^{ps}(3N\log_{2}N\!-\! 3N+4)\!\!$}  & $\!\!U^{ps}(N\log_{2}N \!-\!N+4)\!\!$  & \multicolumn{1}{l|}{$\!\!(U^{ps}+2)(N\log_{2}N \!-\!3N+4)+ 2N\!\!$}  & $\!\!(U^{ps}+2)(N\log_{2}N \!-\!3N+4)+ 2N\!\!$  \\ \hline
\multicolumn{1}{|l|}{CI}     & \multicolumn{1}{l|}{$\!\!U^{ci}(3N\log_{2}N\!-\! 3N+4)\!\!$}  & $\!\!U^{ci}(N\log_{2}N\!-\! 3N+4)\!\!$ & \multicolumn{1}{l|}{$\!\!(2U^{ci}+1)(N\log_{2}N \!-\!3N+4)+ 2N\!\!$} & $\!\!(2U^{ci}+1)(N\log_{2}N \!-\!3N+4)+ 2N\!\!$ \\ \hline
\end{tabular}
\end{table*}

\subsection{Control Signaling Analysis}
\label{subsec:control_signaling_analysis}

The \ac{CIR} separation procedure in~\eqref{eq:fft_extract_cfr} requires explicit coordination between the ISAC BS and the participating \acp{UE}. Specifically, the BS must inform each \ac{UE} of the time-domain sample index \( n_u \) at which to begin extracting its corresponding segment of the estimated \ac{CIR}. Accordingly, the total control signaling overhead \( Q^{i} \) for the proposed APS-ISAC scheme is given by
\begin{equation}
\label{eq:Q_info_prop}
Q^{aps} = U^{aps} \log_2(N).
\end{equation}
In contrast, conventional ISAC schemes incur different levels of control overhead depending on their pilot allocation strategies. For \ac{PS-ISAC} and \ac{CI-ISAC}, the respective control signaling requirements are expressed as
\begin{equation}
\label{eq:Q_info_conv}
Q^{ps} = U^{ps} \log_2\left( \frac{1}{\rho_{cp}} \right), \quad Q^{ci} = U^{ci} \log_2\left( \frac{1}{\rho_p} \right).
\end{equation}
Since \( N_{cp} \) is predefined and shared with all \acp{UE} by the BS, PS-ISAC does not require additional signaling. The only requirement is to convey the phase-shift index~\cite{PS-ISAC}, similar to interleaved subcarrier offset signaling adopted in current standards~\cite{3gpp2025_38211}.

\subsection{Computational Complexity Analysis}
\label{subsec:complexity_analysis}

This subsection analyzes the computational complexity in terms of real additions/subtractions and multiplications/divisions per \ac{OFDM} symbol, focusing on the transmitter side of the ISAC \acp{UE} and the ISAC BS. The receiver side of the ISAC \acp{UE} is excluded from the analysis, as it undergoes no structural modifications.
The complexity of an \( N \)-point FFT is given by \( 3N\log_2N - 3N + 4 \) real additions and \( N\log_2N - 3N + 4 \) real multiplications~\cite{sorensen1986computing}. At the transmitter side of an ISAC \ac{UE}, both APS-ISAC and \ac{PS-ISAC} incur additional computational overhead due to the phase-shift mapping operation in~\eqref{eq:aps_phase_applied}, requiring \( 2N \) real multiplications per \ac{OFDM} symbol to generate \( \bar{\mathbf{X}}_{F,u} \). The subsequent \ac{IFFT} used to obtain the time-domain signal \( \bar{\mathbf{X}}_{T,u} \) in~\eqref{eq:time_domain_transmit_signal} incurs the same complexity across all methods.

At the ISAC BS, FFT operations to obtain \( \mathbf{Y}_F \) in~\eqref{eq:rx_freq_signal} are common to all schemes. Additionally, all scheme requires \( 2N \) real multiplications to compute the normalized frequency response \( \tilde{\mathbf{H}}_F \) in~\eqref{eq:freq_response_estimation}. For CIR estimation, both PS-ISAC and APS-ISAC require only a single \ac{IFFT} to obtain \( \tilde{\mathbf{H}}_T \) in~\eqref{eq:cir_estimation}, regardless of the number of \acp{UE}. In contrast, \ac{CI-ISAC} requires one \ac{IFFT} per \ac{UE} to compute each \( \tilde{\mathbf{H}}_{T,u} \). Furthermore, the computation of each \ac{UE}’s CFR via~\eqref{eq:fft_extract_cfr} necessitates a separate FFT per \ac{UE} in all schemes.
In terms of asymptotic complexity, all methods exhibit \( \mathcal{O}(U^{i} N \log N) \) complexity at the transmitter side. At the BS, APS-ISAC and PS-ISAC achieve \( \mathcal{O}((U^{i} + 2) N \log N) \), while \ac{CI-ISAC} incurs \( \mathcal{O}((2U^{i} + 1) N \log N) \) due to additional per-\ac{UE} \ac{IFFT} operations. A detailed comparison is provided in Table~\ref{table:complexity}.

 \begin{figure}[t]
    \centering
    \vspace*{-0.025\textwidth}
     \hspace*{-0.01\textwidth}
    \includegraphics[width=1.1\linewidth]{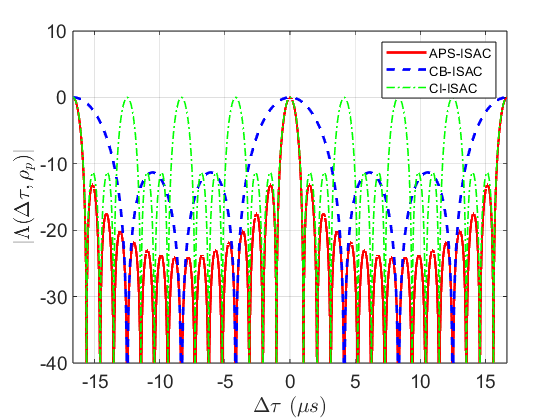}
    \caption{Delay resolution and unambiguous range of conventional and proposed pilot design schemes.}
    \label{fig:res_mur}
\end{figure}

\subsection{Delay and Range Ambiguity \& Resolution Trade-Off}
\label{subsec:tradeoff_delay_resolution}

According to~\cite{xiao2024achieving}, delay sensing performance is characterized using the ambiguity function
\begin{equation}
|\Lambda(\Delta \tau; \rho_p)| = \left| \frac{\sin\left(\pi \Delta f N_p \Delta \tau / \rho_p \right)}{N_p \sin\left(\pi \Delta f \Delta \tau / \rho_p\right)} \right|,
\end{equation}
where \( N_p \) is the number of pilot-bearing subcarriers and \( \Delta \tau \triangleq \tau_{p,u} - \tau_s \) denotes the offset between the actual path delay and the nearest sampled delay grid point. The delay axis is discretized as \( \tau_s = q T_s \), where \( 0 \leq \tau_s < T_{cp} \) and \( T_s = 1/f_s \) is the delay sampling interval.

In \ac{CI-ISAC}, \( \rho_p < 1 \) and \( N_p < N \). In \ac{CB-ISAC}, \( \rho_p = 1 \) but only a subset of subcarriers is utilized, resulting in \( N_p = N \rho_n \). In APS-ISAC, full-band pilot allocation is employed with \( \rho_p = 1 \) and \( N_p = N \). As illustrated in Fig.~\ref{fig:res_mur}, for parameters \( N = 32 \), \( \rho_n = 1/4 \), and \( \rho_p = 1/4 \), the delay resolution corresponds to the half-width of the main lobe of \( |\Lambda(\Delta \tau; \rho_p)| \), which occurs when \( |\pi \Delta f N_p \Delta \tau / \rho_p| = \pi \). This yields the delay and range resolution expressions~\cite{10561589}
\begin{equation}
\tau_{res} = \frac{\rho_p}{N_p \Delta f }, \quad R_{res}^{i} = \frac{ c }{2} \tau_{res} = \frac{\rho_p c }{2 N_p \Delta f }.
\label{eq:delay_res}
\end{equation}
The ambiguity function exhibits grating lobes at
\begin{equation}
\Delta \tau = \frac{\kappa \rho_p}{\Delta f}, \quad \kappa = \pm1, \pm2, \dots, \pm1/\rho_p,
\end{equation}
introducing ambiguity in delay estimation. Accordingly, the unambiguous delay range and the corresponding unambiguous range are given by 
\begin{equation}
\tau_{ua} = \frac{\rho_p}{\Delta f}, \quad R_{ua}^{i} = \frac{c}{2} \tau_{ua} = \frac{N_p c}{2 N \Delta f}.
\label{eq:delay_range}
\end{equation}
Equations~\eqref{eq:delay_res} and~\eqref{eq:delay_range} reveal that \( \Delta f \) governs a fundamental trade-off between delay resolution and unambiguous delay range, as \( \tau_{\mathrm{res}} \propto 1/(N \Delta f) \) and \( \tau_{\mathrm{ua}} \propto 1/\Delta f \). Therefore, selecting \( \Delta f \) involves a trade-off between maximum detectable delay and resolution. As shown in Fig.~\ref{fig:res_mur}, \ac{CB-ISAC} achieves the maximum unambiguous delay range when \( \rho_p = 1 \), yet suffers from degraded delay resolution due to the limited per-\ac{UE} bandwidth. In contrast, \ac{CI-ISAC} improves delay resolution by reducing \( \rho_p \), but this comes at the expense of a reduced unambiguous delay range. In the proposed \ac{APS-ISAC} scheme, full-band pilot allocation enables the system to simultaneously achieve the optimal delay resolution of \( 1/(N \Delta f) \) and the maximum unambiguous delay of \( 1/\Delta f \), thereby overcoming the trade-offs inherent in previous methods. The corresponding range resolution expressions for the three schemes are summarized as follows
\begin{equation}
\label{eq:R_res}
R_{res}^{aps} = R_{res}^{ci} =  \frac{c}{2 N \Delta f}, \quad R_{res}^{cb} =  \frac{c}{2 N \rho_n \Delta f}, 
\end{equation}
while the corresponding maximum unambiguous ranges are
\begin{equation}
\label{eq:R_ua}
R_{ua}^{aps} = R_{ua}^{cb} =  \frac{c}{2 \Delta f}, \quad R_{ua}^{ci} = \frac{c \rho_p}{2 \Delta f}.
\end{equation}
Therefore, APS-ISAC and \ac{CB-ISAC} attain the theoretical maximum unambiguous range by fully utilizing the frequency band. However, \ac{CB-ISAC} experiences degraded range resolution due to underutilization of the available bandwidth per \ac{UE}, while \ac{CI-ISAC} suffers from a reduced unambiguous range due to pilot sparsity.

\begin{figure}
  \centering
     \subfigure[CP size vs number of supported \acp{UE}.]{\includegraphics[width=0.5\textwidth]{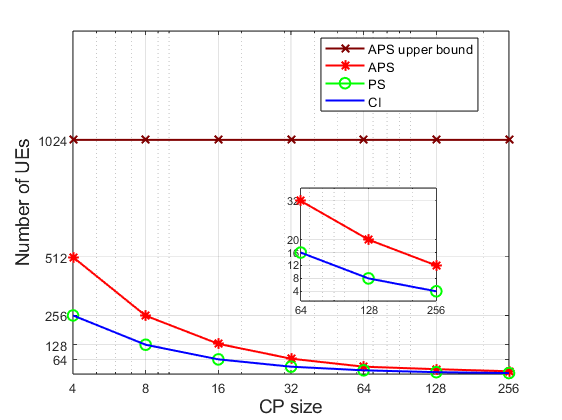}}  
          \subfigure[Number of subcarriers vs number of supported \acp{UE}.]{\includegraphics[width=0.5\textwidth]{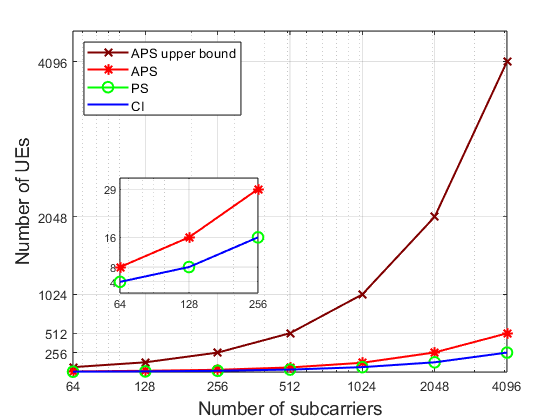}} 
  \caption{SE performance of CI-ISAC, PS-ISAC, and APS-ISAC.}   
      \label{fig:SEs_Figure}
\end{figure}

\begin{figure}
  \centering
     \subfigure[Statistical distributions for the number of channel taps.]{\includegraphics[width=0.5\textwidth]{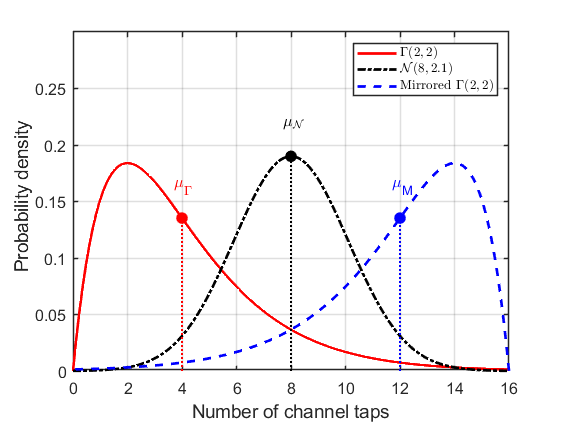}}  
  \subfigure[Number of subcarriers vs number of \acp{UE}.]{\includegraphics[width=0.5\textwidth]{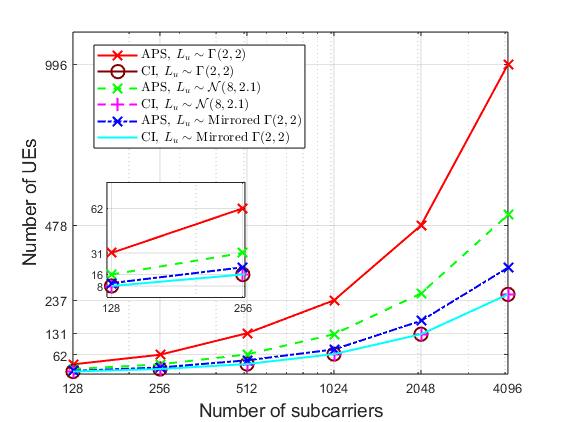}} 
    \caption{SE performances under different distributions for the  channel taps.}   
      \label{fig:SE_vs_distribution}
\end{figure}

\section{Simulation Results}
\label{sec:sim_results}

This section presents comprehensive Monte Carlo simulation results that evaluate the performance of the proposed \ac{APS-ISAC} scheme under various system configurations. The performance metrics include \ac{SE}, \ac{MSE}, computational complexity, and maximum unambiguous range performance. Simulations are conducted over frequency-selective Rayleigh fading communication channels, where the number of taps is randomly selected from the interval \( [1, N_{cp} - 1] \). This setup reflects practical conditions where \( L \leq N_{cp} - 1 \) accounts for delay spread, timing offsets, and guard margins~\cite{tse2005fundamentals}. Unless stated otherwise, pseudo-random pilot sequences are employed with \( N = 256 \) subcarriers. Conventional schemes, including \ac{PS-ISAC} and \ac{CI-ISAC}, operate with a fixed number of \acp{UE}, determined by their respective \( N_{cp} \) and \( \rho_p \) values, as defined  in~\eqref{eq:conv_users}. In contrast, \ac{APS-ISAC} dynamically determines the number of supported \acp{UE} in each simulation trial according to~\eqref{eq:prop_users}, based on the channel tap lengths of individual \acp{UE}.

\begin{figure}[t]
    \centering
    \includegraphics[width=1\linewidth]{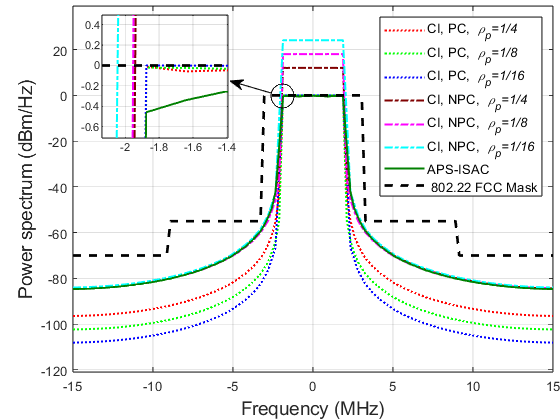}
    \caption{Power spectra comparison of CI-ISAC and APS-ISAC \acp{UE}.}
    \label{fig:PSOFDM}
\end{figure}

\subsection{Spectral Efficiency}

Figs.~\ref{fig:SEs_Figure}(a)--(b) illustrate the \ac{SE} performance of the proposed \ac{APS-ISAC} scheme. In Fig.~\ref{fig:SEs_Figure}(a), the number of subcarriers is fixed at \( N = 1024 \), while Fig.~\ref{fig:SEs_Figure}(b) sets \( N_{cp} = 16 \). In both scenarios, the maximum \( \rho_{cp} \) is set to \( 1/4 \).
Fig.~\ref{fig:SEs_Figure}(a) illustrates the number of supported \acp{UE} as a function of CP size. The upper bound shown for \ac{APS-ISAC} corresponds to the ideal case in which each \ac{UE} experiences a frequency-flat channel. Under this condition, the estimated \acp{CIR} do not overlap at the BS, enabling support for up to \( N \) \acp{UE}. In realistic multi-tap fading channels, \ac{APS-ISAC} consistently outperforms both \ac{PS-ISAC} and \ac{CI-ISAC} across all values of \( N_{cp} \).
This performance gain stems from the statistical characteristics of 
\( L_u \) whose expected value under a normal distribution is
\begin{equation}
    \mathbb{E}[L_u] = \frac{1 + (N_{cp} - 1)}{2} = \frac{N_{cp}}{2}.
\end{equation}
According to \eqref{eq:prop_users}, the expected number of supported \acp{UE} is
\begin{equation}
    \mathbb{E}[U^{aps}] = \frac{N}{\mathbb{E}[L_u]} = \frac{2N}{N_{cp}}.
\end{equation}
Compared to conventional methods in~\eqref{eq:conv_users}, the resulting multiplexing gain factors are
\begin{equation}
    \frac{\mathbb{E}[U^{aps}]}{U^{ci}} = \frac{2N/N_{cp}}{1/\rho_p} = 2, \quad \frac{\mathbb{E}[U^{aps}]}{U^{ps}} = \frac{2N/N_{cp}}{1/\rho_{cp}} = 2.
\end{equation}
Therefore, \ac{APS-ISAC} supports approximately twice as many \acp{UE} as the benchmark schemes when \( L_u \sim \mathcal{N}(N_{cp}/2, \sigma^2) \). Since \( \rho_p = \rho_{cp} \) is assumed, \ac{PS-ISAC} and \ac{CI-ISAC} yield identical \ac{SE} performance.
Fig.~\ref{fig:SEs_Figure}(b) demonstrates the effect of increasing \( N \) on the number of supported \acp{UE}. As \( N \) increases, \ac{APS-ISAC} scales efficiently, approaching the theoretical upper bound of \( N \) under ideal conditions. The relative gain over conventional schemes remains stable, confirming the scalability of \ac{APS-ISAC} for large-bandwidth applications.

We now evaluate the impact of different statistical models for the number of channel taps \( L \) on the performance of the proposed \ac{APS-ISAC} scheme. Fig.~\ref{fig:SE_vs_distribution} compares \ac{APS-ISAC} with benchmark \ac{PS-ISAC} and \ac{CI-ISAC} schemes under three distinct channel tap generation models.
Fig.~\ref{fig:SE_vs_distribution}(a) displays the \acp{PDF} of Gamma, Normal, and mirrored Gamma distributions, each characterizing a different multipath profile. For illustrative purposes, different delay spread conditions are modeled using distinct statistical distributions. A Gamma distribution with parameters \( \Gamma(2,2) \) captures short delay spreads with an expected number of taps \( \mu_\Gamma = 4 \), whereas a Normal distribution \( \mathcal{N}(8, 2.1) \) characterizes moderate spreads with \( \mu_N = 8 \). To represent longer delay profiles, a mirrored Gamma distribution is employed with an expected value \( \mu_M = 12 \). All distributions are truncated within $[0, N_{cp}]$. These parameters are chosen to emulate distinct delay characteristics and can be replaced with alternative values depending on the propagation environment.

Fig.~\ref{fig:SE_vs_distribution}(b) illustrates the number of supported \acp{UE} as a function of \( N \) for each channel model. Across all configurations, \ac{APS-ISAC} consistently outperforms the benchmark schemes. Notably, under the Gamma distribution, \ac{APS-ISAC} supports nearly 1000 \acp{UE} for \( N = 4096 \), while \ac{PS-ISAC} and \ac{CI-ISAC} are limited to approximately 250. As the expected number of taps increases (e.g., under the mirrored Gamma case), the performance of \ac{APS-ISAC} gradually declines due to the reduced number of \acp{UE} that can be accommodated within \( N \), although it still maintains a substantial performance advantage.
This gain stems from the adaptive phase-shift mechanism in \ac{APS-ISAC}, which dynamically accounts for the \ac{UE}-specific maximum number of channel taps to prevent pilot collisions. In contrast, \ac{PS-ISAC} and \ac{CI-ISAC} lack such delay-aware design, resulting in degraded scalability in dense multipath environments.

\subsection{Power Spectrum and MSE}

The spectral characteristics of \ac{APS-ISAC} and \ac{CI-ISAC} schemes are compared in Fig.~\ref{fig:PSOFDM} under both \ac{PC} and non-\ac{PC} scenarios, across different pilot ratios.
Since \ac{PS-ISAC} exhibits an identical power spectral density to \ac{APS-ISAC}, its spectrum is omitted for visual clarity.
In the \ac{PC} setting, each pilot subcarrier is normalized to unit power to ensure compliance with standardized spectral constraints. In the non-\ac{PC} case, each pilot subcarrier is scaled by \( \sqrt{1/\rho_p} \) to maintain equal total transmit power relative to the \( \rho_p = 1 \) benchmark.
While the \ac{PC} configuration represents a practical, standard-compliant scenario, the non-\ac{PC} case highlights the spectral implications of maintaining equal transmit energy across schemes. As shown in Fig.~\ref{fig:PSOFDM}, the non-\ac{PC} configuration for \ac{CI-ISAC} exhibits substantial spectral leakage, exceeding the IEEE 802.22 FCC emission mask~\cite{tom2013mask}, particularly around the mainlobe. In contrast, both \ac{APS-ISAC} and \ac{CI-ISAC} under \ac{PC} remain within regulatory bounds.
\ac{APS-ISAC} achieves significantly improved spectral containment, with sidelobes more closely adhering to the FCC mask compared to the benchmark methods. This result demonstrates that \ac{APS-ISAC} provides enhanced spectral shaping capabilities while remaining compliant with emission regulations.

\begin{figure}
  \centering
     \subfigure[Frequency flat channel.]{\includegraphics[width=0.5\textwidth]{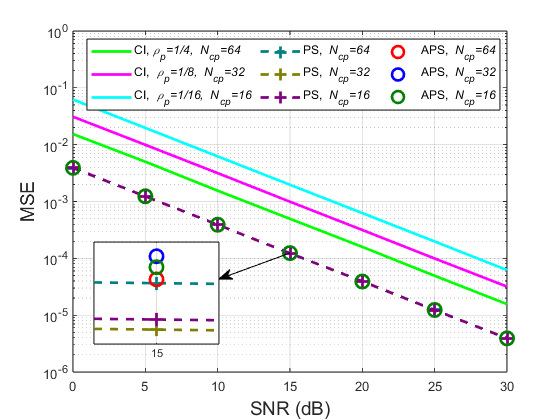}}  
     \subfigure[Frequency-selective channel.]{\includegraphics[width=0.5\textwidth]{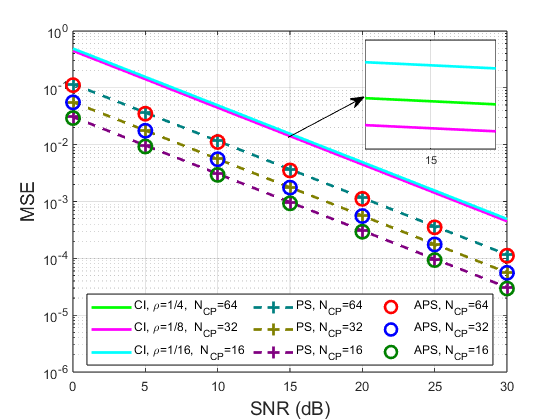}} 
  \caption{MSE performance of CI-ISAC, PS-ISAC, and APS-ISAC under power constraints.}   
      \label{fig:MSEs_under_PC_All_transmitters}
\end{figure}

\begin{figure}
  \centering
     \subfigure[Frequency flat channel.]{\includegraphics[width=0.5\textwidth]{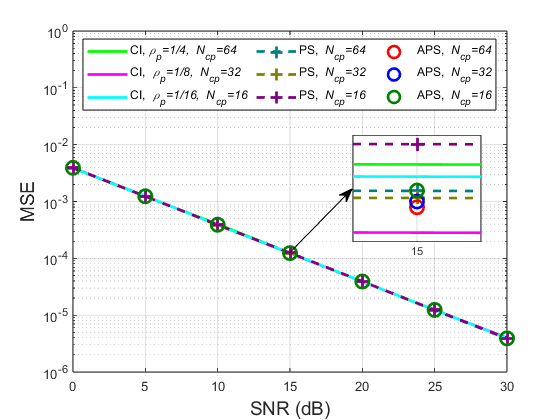}}
     \subfigure[Frequency-selective channel.]{\includegraphics[width=0.5\textwidth]{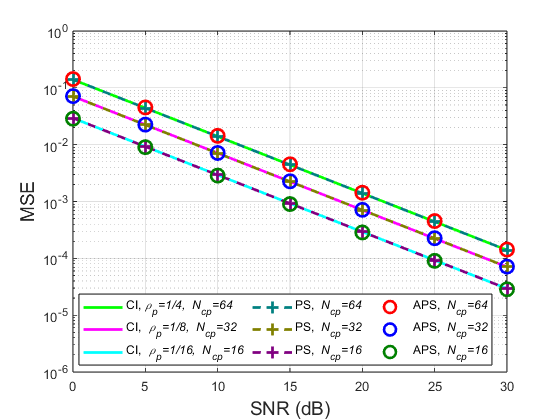}} 
  \caption{MSE performance of CI-ISAC, PS-ISAC, and APS-ISAC without power constraints.}   
      \label{fig:MSEs_without_PC_All_transmitters}
\end{figure}

Fig.~\ref{fig:MSEs_under_PC_All_transmitters} presents the \ac{MSE} performance of \ac{APS-ISAC} and the benchmark schemes under power-constrained operation. Across all pilot ratios, \ac{APS-ISAC} consistently outperforms \ac{CI-ISAC} while achieving nearly identical performance to \ac{PS-ISAC}, despite supporting a significantly larger number of \acp{UE}. This robustness arises from the ability of \ac{APS-ISAC} to allocate all subcarriers to each \ac{UE}. In contrast, \ac{CI-ISAC} exhibits degraded \ac{MSE} as the number of \acp{UE} increases relative to the fixed pilot resources. The \ac{MSE} is calculated as $\frac{1}{U^i} \sum_{u=1}^{U^i} \mathbb{E}[(\mathbf{H}_{F,u}-\Tilde{\mathbf{H}}_{F, u})^2]$.
Fig.~\ref{fig:MSEs_under_PC_All_transmitters}(a) reports results under frequency-flat channel conditions. In this setting, the noise contribution remains constant since only a single tap is estimated per \ac{UE}. As the total pilot power allocated to each \ac{UE} decreases with \( \rho_p \), the \ac{MSE} of \ac{CI-ISAC} worsens with decreasing \( \rho_p \). On the other hand, \ac{APS-ISAC} maintains constant pilot power per \ac{UE} by using the entire subcarrier set (\( \rho_p = 1 \)), yielding stable \ac{MSE} performance regardless of the number of \acp{UE}.
In Fig.~\ref{fig:MSEs_under_PC_All_transmitters}(b), \ac{APS-ISAC} demonstrates increasing \ac{MSE} with larger \( N_{cp} \), as longer \acp{CIR} introduce more taps to estimate, thereby accumulating more noise. For \ac{CI-ISAC}, although the number of taps also increases with \( N_{cp} \), the power per pilot subcarrier diminishes due to unit-power normalization, resulting in convergence of the \ac{MSE} curves across different \( N_{cp} \) values.
The key trade-off in \ac{APS-ISAC} lies in its higher pilot power consumption per \ac{UE} compared to \ac{CI-ISAC}. Nevertheless, in contrast to \ac{PS-ISAC}, it supports a substantially larger number of \acp{UE} while maintaining comparable \ac{MSE} performance. As illustrated in Figs.~\ref{fig:MSEs_under_PC_All_transmitters}(a)--(b), reducing the expected number of channel taps improves estimation accuracy, underscoring the impact of multipath richness in channel environments.

Fig.~\ref{fig:MSEs_without_PC_All_transmitters} depicts \ac{MSE} results under relaxed power constraints, where each scheme scales its pilot energy proportionally to the number of occupied subcarriers. Under this configuration, \ac{APS-ISAC}, \ac{PS-ISAC}, and \ac{CI-ISAC} exhibit nearly identical \ac{MSE} performance for all \( N_{cp} \) values.
In Fig.~\ref{fig:MSEs_without_PC_All_transmitters}(a), the \ac{MSE} remains constant across \( N_{cp} \) due to the frequency-flat channel model, where only a single noise sample affects the estimation. Conversely, Fig.~\ref{fig:MSEs_without_PC_All_transmitters}(b) reveals that the \ac{MSE} increases with \( N_{cp} \) for all schemes, reflecting the growing number of taps in frequency-selective channels. These findings validate that the performance gaps observed in Fig.~\ref{fig:MSEs_under_PC_All_transmitters} primarily originate from differences in pilot power allocation strategies rather than from estimation capability.

\begin{table*}[]
\centering
\caption{Comparison of required control information parameters and computational complexity for APS-ISAC, PS-ISAC, and CI-ISAC.} 
\label{tab:control_info_comp}
\begin{tabular}{clllcc|cccc|}
\cline{7-10}
\multicolumn{1}{l}{}                            &                               &                                  &                          & \multicolumn{1}{l}{}     & \multicolumn{1}{l|}{} & \multicolumn{4}{c|}{\textbf{Computational Complexity}}                                                                                                          \\ \cline{2-10} 
\multicolumn{1}{c|}{}                           & \multicolumn{5}{c|}{\textbf{Control Signaling}}                                                                                              & \multicolumn{2}{c|}{Transmitter Side of ISAC \Ac{UE}}                                  & \multicolumn{2}{c|}{ISAC BS}                             \\ \hline
\multicolumn{1}{|c|}{Method}                    & \multicolumn{1}{l|}{$\rho_p$} & \multicolumn{1}{l|}{$\rho_{cp}$} & \multicolumn{1}{l|}{$Q^{i}$} & \multicolumn{1}{c|}{$U^{i}$} & $\eta$                & \multicolumn{1}{c|}{Addition--Per UE} & \multicolumn{1}{c|}{Multiplication--Per UE} & \multicolumn{1}{c|}{Addition-Per UE} & Multiplication-Per UE \\ \hline
\multicolumn{1}{|c|}{\multirow{3}{*}{APS-ISAC}} & \multicolumn{1}{l|}{1}        & \multicolumn{1}{l|}{1/4}         & \multicolumn{1}{l|}{64}  & \multicolumn{1}{c|}{8}   & \multirow{3}{*}{8}    & \multicolumn{1}{c|}{43040--5380}       & \multicolumn{1}{c|}{14368--1796}             & \multicolumn{1}{c|}{53800--6725}      & 13352--1669            \\ \cline{2-5} \cline{7-10} 
\multicolumn{1}{|c|}{}                          & \multicolumn{1}{l|}{1}        & \multicolumn{1}{l|}{1/8}         & \multicolumn{1}{l|}{128} & \multicolumn{1}{c|}{16}  &                       & \multicolumn{1}{c|}{86080--5380}       & \multicolumn{1}{c|}{28736--1796}             & \multicolumn{1}{c|}{96840--6053}      & 23624--1477            \\ \cline{2-5} \cline{7-10} 
\multicolumn{1}{|c|}{}                          & \multicolumn{1}{l|}{1}        & \multicolumn{1}{l|}{1/16}        & \multicolumn{1}{l|}{256} & \multicolumn{1}{c|}{32}  &                       & \multicolumn{1}{c|}{172160--5380}      & \multicolumn{1}{c|}{57472--1796}             & \multicolumn{1}{c|}{182920--5717}     & 44168--1381            \\ \hline
\multicolumn{1}{|c|}{\multirow{3}{*}{PS-ISAC}}  & \multicolumn{1}{l|}{1}        & \multicolumn{1}{l|}{1/4}         & \multicolumn{1}{l|}{8}   & \multicolumn{1}{c|}{4}   & 2                     & \multicolumn{1}{c|}{21520--5380}       & \multicolumn{1}{c|}{7184--1796}              & \multicolumn{1}{c|}{32280--8070}      & 8216--2054             \\ \cline{2-10} 
\multicolumn{1}{|c|}{}                          & \multicolumn{1}{l|}{1}        & \multicolumn{1}{l|}{1/8}         & \multicolumn{1}{l|}{24}  & \multicolumn{1}{c|}{8}   & 3                     & \multicolumn{1}{c|}{43040--5380}       & \multicolumn{1}{c|}{14368--1796}             & \multicolumn{1}{c|}{53800--6725}      & 13352--1669            \\ \cline{2-10} 
\multicolumn{1}{|c|}{}                          & \multicolumn{1}{l|}{1}        & \multicolumn{1}{l|}{1/16}        & \multicolumn{1}{l|}{64}  & \multicolumn{1}{c|}{16}  & 4                     & \multicolumn{1}{c|}{86080--5380}       & \multicolumn{1}{c|}{28736--1796}             & \multicolumn{1}{c|}{96840--6053}      & 23624--1477            \\ \hline
\multicolumn{1}{|c|}{\multirow{3}{*}{CI--ISAC}} & \multicolumn{1}{l|}{1/4}      & \multicolumn{1}{l|}{1/4}         & \multicolumn{1}{l|}{8}   & \multicolumn{1}{c|}{4}   & 2                     & \multicolumn{1}{c|}{21520--5380}       & \multicolumn{1}{c|}{5136--1284}              & \multicolumn{1}{c|}{48420--12105}     & 12068--3017            \\ \cline{2-10} 
\multicolumn{1}{|c|}{}                          & \multicolumn{1}{l|}{1/8}      & \multicolumn{1}{l|}{1/8}         & \multicolumn{1}{l|}{24}  & \multicolumn{1}{c|}{8}   & 3                     & \multicolumn{1}{c|}{43040--5380}       & \multicolumn{1}{c|}{10272--1284}             & \multicolumn{1}{c|}{91460--11433}     & 22340--2793            \\ \cline{2-10} 
\multicolumn{1}{|c|}{}                          & \multicolumn{1}{l|}{1/16}     & \multicolumn{1}{l|}{1/16}        & \multicolumn{1}{l|}{64}  & \multicolumn{1}{c|}{16}  & 4                     & \multicolumn{1}{c|}{86080--5380}       & \multicolumn{1}{c|}{20544--1284}             & \multicolumn{1}{c|}{177540--11097}    & 42884--2681            \\ \hline
\end{tabular}
\end{table*}

\subsection{Control Signaling and Computational Complexity}
\label{subsec:complexity_control}

Table~\ref{tab:control_info_comp} summarizes the control signaling and computational complexity characteristics of the proposed and benchmark schemes as functions of \( \rho_p \) and \( \rho_{cp} \). 
For control signaling, the expected overhead depends on the number of multiplexed \acp{UE} and their coordination requirements. The signaling overhead for \ac{APS-ISAC} is modeled by~\eqref{eq:Q_info_prop}, while \ac{PS-ISAC} and \ac{CI-ISAC} follow~\eqref{eq:Q_info_conv}. Although \ac{APS-ISAC} introduces slightly higher control overhead relative  to the benchmarks schemes, the increase is moderate and justified by the significantly improved scalability in \ac{UE} multiplexing and SE. Furthermore, while \ac{PS-ISAC} and \ac{CI-ISAC} experience linearly growing control load with the number of \acp{UE}, \ac{APS-ISAC} maintains a stable per-\ac{UE} signaling burden \( \eta \) due to its time sample coordination mechanism.

The computational complexity analysis is centered on the transmitter-side operations of the ISAC \ac{UE}, as the receiver-side processing remains identical across all schemes, as previously discussed in Subsection \ref{subsec:complexity_analysis}. \ac{APS-ISAC} and \ac{PS-ISAC} incur the same per-\ac{UE} computational cost, requiring 5380 additions and 1796 multiplications. In contrast, \ac{CI-ISAC} performs the same number of additions but only 1284 multiplications due to the absence of adaptive phase-shift operations. The additional operations in \ac{APS-ISAC} involve element-wise multiplications and remain practically negligible in complexity, as they do not scale with \( \mathcal{O}(N \log N) \).

A major advantage of \ac{APS-ISAC} arises at the BS side: \ac{CI-ISAC} requires one IFFT per \ac{UE}, resulting in a BS-side complexity that scales linearly with \( U^{ci} \). In contrast, \ac{APS-ISAC} uses a single IFFT and applies circular shift-based post-processing to isolate each \ac{UE}’s CIR, significantly reducing the computational load. For large \( U^{i} \), \ac{APS-ISAC} achieves nearly half the per-\ac{UE} BS-side complexity of \ac{CI-ISAC}. As the system scales, \ac{CI-ISAC}'s BS becomes a bottleneck with \( \mathcal{O}((2U^{i} + 1) N \log N) \) complexity, whereas \ac{APS-ISAC} maintains a more balanced load at \( \mathcal{O}((U^{i} + 2) N \log N) \). When compared to \ac{PS-ISAC}, the BS-side computational complexity remains the same when serving an equal number of \acp{UE}. However, since APS-ISAC supports a higher degree of \ac{UE} multiplexing under same time-frequency resources, the per-\ac{UE} complexity at the BS is effectively reduced in APS-ISAC compared to PS-ISAC.

\subsection{Range Unambiguity and Resolution}
\label{subsec:sim_MUR}

This subsection evaluates the sensing performance of APS-ISAC, CI-ISAC, and CB-ISAC in terms of maximum unambiguous range and range resolution under a monostatic uplink OFDM-based ISAC scenario. Simulations are conducted using the parameters listed in Table~\ref{tab:sim_params}, where a single \ac{UE} performs radar-based sensing to isolate the impact of pilot structure on estimation performance. Radar targets with varying ranges and velocities are positioned in close proximity to the sensing \ac{UE}, enabling evaluation of delay-Doppler resolution. Accordingly, interference from other \acp{UE} is neglected. 

\begin{figure}
    \centering
    \subfigure[APS-ISAC with $\rho_p=1$ \& $N=128$.]{
    \makebox[\linewidth][l]{\hspace{-0.06\linewidth}\includegraphics[width=1.15\linewidth]{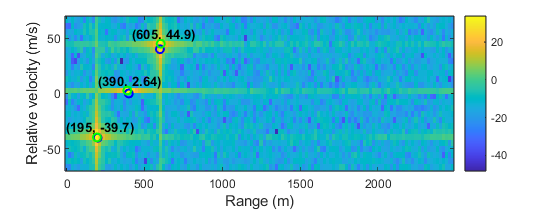}}} \\
        \subfigure[CI-ISAC with $\rho_p=1/4$ \& $N=128$.]{
    \makebox[\linewidth][l]{\hspace{-0.06\linewidth}\includegraphics[width=1.15\linewidth]{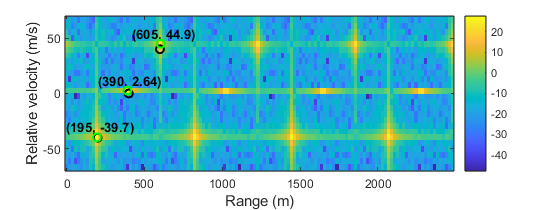}}} \\
        \subfigure[CI-ISAC with $\rho_p=1/8$ \& $N=128$.]{
    \makebox[\linewidth][l]{\hspace{-0.06\linewidth}\includegraphics[width=1.15\linewidth]{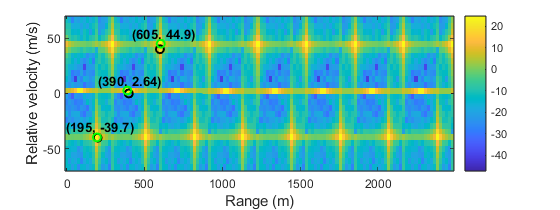}}} \\
        \subfigure[CB-ISAC with $\rho_p=1$ \& $\rho_n=1/4$.]{
    \makebox[\linewidth][l]{\hspace{-0.06\linewidth}\includegraphics[width=1.15\linewidth]{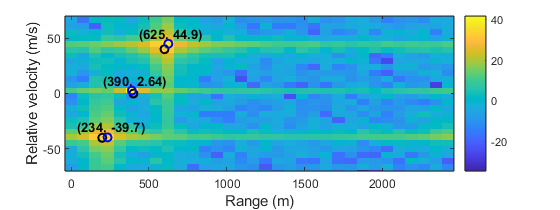}}} \\
        \subfigure[CB-ISAC with $\rho_p=1$ \& $\rho_n=1/8$.]{
    \makebox[\linewidth][l]{\hspace{-0.06\linewidth}\includegraphics[width=1.15\linewidth]{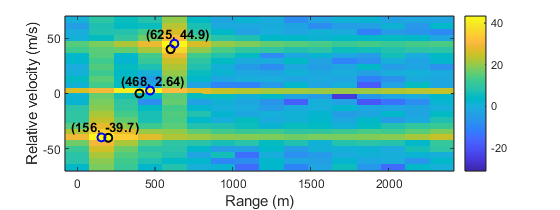}}} \\
      \caption{Range–velocity plots of the targets for maximum unambiguous range comparison without PC.}   
          \label{fig:MURs_without_PC}
\end{figure}

For CI-ISAC, pilot ratios \( \rho_p \in \{1/4, 1/8\} \) are considered, while CB-ISAC is evaluated with subcarrier configurations \( \rho_n \in \{1/4, 1/8\} \). The sensing \( \mathrm{UE}_{U^{i}} \) processes its own reflected signals, and the resulting range–velocity maps are shown in Figs.~\ref{fig:MURs_without_PC}(a)--(e), where true target positions are marked with black circles and estimated positions are indicated using colored markers consistent with the legend in Fig.~\ref{fig:res_mur}. PS-ISAC and APS-ISAC employs full pilot reuse at all subcarriers (\( \rho_p = \rho_n = 1 \)) and adaptive phase shifts across all subcarriers. As detailed in Section~\ref{subsec:tradeoff_delay_resolution}, this configuration yields the finest range resolution according to~\eqref{eq:R_res}
\begin{equation}
\begin{aligned}
& R_{res}^{\mathrm{aps}} = R_{res}^{\mathrm{ci}} = \frac{c}{2 \cdot 128 \cdot 60 \cdot 10^3} \approx 19.52~\text{m}, \\
& R_{res}^{\mathrm{cb}} =
\begin{cases}
\displaystyle \frac{c}{2 \cdot 128 \cdot \sfrac{1}{4} \cdot 60 \cdot 10^3} \approx 78.07~\text{m}, & \rho_n = \frac{1}{4} \\
\displaystyle \frac{c}{2 \cdot 128 \cdot \sfrac{1}{8} \cdot 60 \cdot 10^3} \approx 156.14~\text{m}, & \rho_n = \frac{1}{8}.
\end{cases}
\end{aligned}
\end{equation}
The corresponding maximum unambiguous range, as defined in~\eqref{eq:R_ua}, is given by
\begin{equation}
\begin{aligned}
& R_{ua}^{\mathrm{aps}} = R_{ua}^{\mathrm{cb}} = \frac{c}{2 \Delta f} = \frac{c}{2 \cdot 60 \cdot 10^3} \approx 2498~\text{m}, \\
& R_{ua}^{\mathrm{ci}} =
\begin{cases}
\displaystyle \frac{c \cdot \sfrac{1}{4}}{2 \cdot 60 \cdot 10^3} \approx 624~\text{m}, & \rho_p = \frac{1}{4} \\
\displaystyle \frac{c \cdot \sfrac{1}{8}}{2 \cdot 60 \cdot 10^3} \approx 312~\text{m}, & \rho_p = \frac{1}{8}.
\end{cases}
\end{aligned}
\end{equation}

\begin{table}[t]
\centering
\caption{Radar-based sensing simulation parameters.}
\label{tab:sim_params}
\begin{tabular}{|c|l|l|}
\hline
\textbf{Symbol} & \multicolumn{1}{c|}{\textbf{Parameter}} & \multicolumn{1}{c|}{\textbf{Value}} \\ \hline
$N$             & Number of subcarriers                   & $128$                               \\ \hline
$M$             & Number of OFDM symbols                  & $64$                                \\ \hline
$f_c$           & Carrier frequency (GHz)                 & $24$                                \\ \hline
$f_s$           & Sampling frequency (MHz)                & $7.68$                              \\ \hline
$\Delta f$      & Subcarrier spacing (kHz)                & $60$                                \\ \hline
$T$             & OFDM symbol duration $(\mu s)$          & $\approx 16.67$                     \\ \hline
$T_{cp}$        & CP duration $(\mu s)$                   & $\approx 4.1667$                    \\ \hline
$R_p$           & Target ranges (m)                       & $200$, $400$, $600$              \\ \hline
$v_p$           & Target velocities (m/s)                 & $-40$, $0$, $+40$                   \\ \hline
$\rho_p$        & Pilot ratios in \ac{CI-ISAC}                           & $1/4, 1/8$                          \\ \hline
$\rho_n$        & Subcarrier ratio in \ac{CB-ISAC}       & $1/4, 1/8$                              \\ \hline
SNR             & Signal-to-noise ratio (dB)              & $10$                                \\ \hline
\end{tabular}
\end{table}

In Fig.~\ref{fig:MURs_without_PC}(a), APS-ISAC demonstrates accurate and unambiguous detection of all targets. CI-ISAC uses interleaved subcarrier mappings that effectively thin the spectrum as \( \rho_p \) decreases. As shown in Figs.~\ref{fig:MURs_without_PC}(b)--(c), this reduces the unambiguous range due to increased spacing between pilot subcarriers. The results in the appearance of periodic ambiguity peaks, whose density doubles for each halving of \( \rho_p \).
CB-ISAC, employing localized orthogonal pilot blocks, achieves a similar unambiguous range to APS-ISAC, as depicted in Figs.~\ref{fig:MURs_without_PC}(d)--(e). However, its range resolution is substantially worse, evidenced by broader main lobes and degraded peak localization. This degradation arises due to underutilization of the available frequency band. Quantitatively, the range estimation performance is evaluated using the \ac{MSE}, defined as \( \mathrm{MSE}_{R,U^{i}} = \frac{1}{P} \sum_{p=1}^{P} \mathbb{E}[(R_{p,U^{i}} - \hat{R}_{p,U^{i}})^2] \), is observed to be \( \mathrm{MSE}_{R,U^{\mathrm{aps}}} = \mathrm{MSE}_{R,U^{\mathrm{ci}}} = 50 \) for both APS-ISAC and CI-ISAC. In contrast, CB-ISAC yields substantially higher errors of \( \mathrm{MSE}_{R,U^{\mathrm{cb}}} = 734 \) and 3129 for subcarrier densities \( \rho_n = 1/4 \) and \( \rho_n = 1/8 \), respectively. 

Overall, the results corroborate the theoretical analysis presented in Section~\ref{subsec:tradeoff_delay_resolution}. APS-ISAC simultaneously achieves the maximum unambiguous range and the finest range resolution by utilizing full-bandwidth pilot reuse in conjunction with adaptive phase shifts. While CI-ISAC maintains high resolution, it suffers from reduced unambiguous range at lower pilot densities. In contrast, CB-ISAC preserves the unambiguous range but experiences significant degradation in range resolution. PS-ISAC is excluded from this comparison, as it operates under the same pilot ratio framework as APS-ISAC and exhibits identical performance in terms of range ambiguity and resolution. These findings emphasize the importance of jointly optimizing pilot allocation and subcarrier structures in ISAC systems operating under bandwidth constraints.

\section{Concluding Remarks}
\label{sec:conclusion}

This paper proposed \ac{APS-ISAC}, a novel pilot design framework for uplink multiple access in ISAC systems which employs adaptive phase shifts for enabling time-domain separation of multi-\ac{UE} \acp{CIR}. By tailoring phase shifts to the maximum excess delay rather than limiting them by the \ac{CP} length, \ac{APS-ISAC} extends the \ac{PS-ISAC} scheme to support scalable \ac{UE} multiplexing.
Theoretical analysis showed that \ac{APS-ISAC} achieves high range resolution and maximum unambiguous range by utilizing the full set of available subcarriers, while enabling fully overlapping pilot reuse across \acp{UE}. Additionally, the scheme facilitates efficient UE multiplexing with negligible control signaling and reduced complexity. Notably, \ac{APS-ISAC} eliminates the need for per-\ac{UE} IFFT operations at the BS, significantly lowering BS-side complexity relative to \ac{CI-ISAC}, and reduces per-\ac{UE} processing overhead compared to \ac{PS-ISAC}.
The performance benefits of APS-ISAC were validated through extensive simulations. The proposed method achieves higher \ac{SE}, supporting nearly twice the number of multiplexed \acp{UE} on average relative to \ac{CI-ISAC} and \ac{PS-ISAC}. It achieves lower estimation \ac{MSE} under power constraints than \ac{CI-ISAC}, maximizes unambiguous range over \ac{CI-ISAC}; improves sensing resolution over \ac{CB-ISAC}, and outperforms both \ac{CI-ISAC} and \ac{PS-ISAC} in computational efficiency. These results demonstrate that APS-ISAC offers a scalable and spectrally efficient pilot design paradigm for uplink ISAC systems in IoT and V2X applications.

Future work may extend the framework to address mutual radar interference among \acp{UE}, and the joint optimization of sensing and communication under different data transmission scenarios.



\vfill

\end{document}